\documentclass[aps,manuscript,showpacs,showkeys]{revtex4}
\usepackage{graphicx}% Include figure files
\usepackage{epsfig}  % Include figure files
\usepackage{epsf}    % Include figure files
\usepackage{dcolumn}% Align table columns on decimal point
\usepackage{bm}% bold math
\usepackage{dcolumn}% Align table columns on decimal point
\usepackage{textcomp}% Align table columns on decimal point
\usepackage[tbtags]{amsmath}
\usepackage{amsfonts}
\usepackage{float}
\usepackage[justification=justified,singlelinecheck=false]{caption}

\newcommand{\be}{\begin{equation}}
\newcommand{\ee}{\end{equation}}
\newcommand{\bea}{\begin{eqnarray}}
\newcommand{\eea}{\end{eqnarray}}

\newcommand{\tjk}{{\theta_{12}}}
\newcommand{\tjl}{{\theta_{13}}}
\newcommand{\tkl}{{\theta_{23}}}
\newcommand{\tkllo}{{\theta_{23}^{LO}}}
\newcommand{\tklho}{{\theta_{23}^{HO}}}
\newcommand{\mlj}{{\Delta_{31}}}
\newcommand{\mkj}{{\Delta_{21}}}
\newcommand{\mlk}{\Delta_{32}}
\newcommand{\mm}{\Delta_{\mu\mu}}

\newcommand{\hatmlj}{\hat{\Delta}_{31}}
\newcommand{\hatmkj}{\hat{\Delta}_{21}}
\newcommand{\hatmlk}{\hat{\Delta}_{32}}
\newcommand{\hatmm}{\hat{\Delta}_{\mu\mu}}

\newcommand{\dcp}{{\delta_{CP}}}
\newcommand{\nova}{{NO$\nu$A }}

\newcommand{\pmumu}{P_{\mu\mu}}

\newcommand{\xx}{{\chi^2}}
\newcommand{\mcomp}{$\mu$-complementary }

\newcommand{\sskjl}{\sin^2 2\theta_{13}}
\newcommand{\skl}{\sin\theta_{23}}

\newcommand{\sskkl}{\sin^2 2\theta_{23}}
\newcommand{\tm}{\theta_{\mu\mu}}
\newcommand{\tmlo}{\theta_{\mu\mu}^{LO}}
\newcommand{\tmho}{\theta_{\mu\mu}^{HO}}

\newcommand{\sskm}{\sin^2 2\theta_{\mu\mu}}

\begin{document}
% \ifpdf
%\DeclareGraphicsExtensions{.pdf,.jpg,.mps,.png}
% \else
%\DeclareGraphicsExtensions{.eps,.ps}
% \fi

%\preprint{}

\title{Effect of non-zero $\theta_{13}$ on the measurement of $\theta_{23}$}


%%%%%%%%%%%%%%%%%%%%%%%%%%%%%%%%%%%%%%%%%%%%%%%%%%%%%
\author{Sushant K. Raut}
\email[Email Address: ]{sushant@prl.res.in}
\affiliation{ Physical Research Laboratory, Ahmedabad, India}

%\collaboration can be followed by \email, \homepage, 
%\thanks as well.
\date{\today}
%%%%%%%%%%%%%%%%%%%% abstract %%%%%%%%%%%%%%%%%%%%%%%%%%%%%%%%%%%%%%%%%%
\begin{abstract}
The moderately large measured value of $\tjl$ signals a departure from the approximate 
two-flavour oscillation framework. As a consequence, the relation between the value of 
$\tkl$ in nature, and the mixing angle measured in $\nu_\mu$ disappearance experiments 
is non-trivial. In this paper, we calculate this relation analytically. We also derive the 
correct conversion between degenerate values of $\tkl$ in the two octants. Through 
simulations of a $\nu_\mu$ disappearance experiment, we show that there are 
observable consequences of not using the correct relation in calculating oscillation 
probabilities. These include a wrong best-fit 
value for $\tkl$, and spurious sensitivity to the octant of $\tkl$. 
\end{abstract}
%%%%%%%%%%%%%%%%%%%%%%%%%%%%%%%%%%%%%%%%%%%%%%%%%%%%%%%%%%%%%%%%%%%%%%%%%%
%\maketitle must follow title, authors, abstract, 
%\pacs, and \keywords
\maketitle

\section{Introduction}

Neutrino oscillation physics has entered an era of precision measurements. 
Since 2011, the reactor experiments Daya Bay \cite{dayabay}, Double Chooz 
\cite{dchooz1,dchooz2} and RENO \cite{reno} 
and superbeam experiments MINOS \cite{minos} and T2K \cite{t2k} have measured 
a non-zero value of $\tjl$ \cite{dayabay_t13,dchooz_dec2011,reno_t13}. 
Analyses of world neutrino data \cite{fogli2012,tortola2012,schwetz_nufit} have 
given us the value $\sskjl \simeq 0.1$, which is moderately large. Daya Bay 
has measured $\sskjl = 0.089 \pm 0.011$ \cite{dayabay_NF12}, which is the most 
precise measurement till date. These measurements have established that $\tjl$ 
is non-zero at more than $5\sigma$ confidence level. 

The solar mixing angle $\tjk$ and mass-squared difference $\mkj$ 
($\Delta_{ij}=m_i^2 - m_j^2$) have been 
measured very accurately by SNO \cite{sno} and KamLAND \cite{kamland}, respectively. 
Their current best-fit values are $\sin^2\tjk = 0.32 \pm 0.017$ and 
$\mkj = (7.62 \pm 0.2) \times 10^{-5} \textrm{eV}^2$ \cite{fogli2012}. 
MINOS \cite{minos} has measured 
the mixing angle $\tm$ and mass-squared difference $\mm$
with a precision of a few percent. We use the subscript $\mu\mu$ to indicate that 
these parameters are measured from observations of muon neutrino disappearance. 
The values of these parameters from MINOS are $\sskm > 0.90$ ($90\%$ C.L.) and 
$|\mm| = (2.32 \pm 0.1) \times 10^{-3} \textrm{eV}^2$ \cite{minos_2011}. 
In the two-flavour oscillation scenario (neglecting the small parameters 
$\mkj$ and $\tjl$), the muon neutrino survival probability $\pmumu$ depends on 
$\sskm=\sskkl$ and $|\mm|=|\mlj|$. Because this probability depends 
on the magnitude but not 
the sign of $|\mlj|$, we cannot determine the neutrino mass ordering or hierarchy.
Moreover, this function is symmetric about $\tkl=45^\circ$, 
which gives 
rise to the octant degeneracy. Thus, 
the currently unknown parameters in standard three-flavour neutrino oscillation 
physics are - (a) the 
neutrino mass hierarchy (normal hierarchy (NH): $\mm>0$ or inverted hierarchy (IH): 
$\mm<0$), (b) the CP-violating phase $\dcp$ and (c) the octant of $\tm$ 
(lower octant (LO): $\tm<45^\circ$ or higher octant (HO): $\tm>45^\circ$). 

It is 
possible for the mass hierarchy to be determined by the upcoming experiment \nova 
itself, if the value of $\dcp$ in nature is in the favourable range \cite{nova_tdr}. 
Combined data from multiple experiments \cite{degeneracy1,hubercpv,novat2k}, 
experiments with longer baselines \cite{lbne_interim2010,laguna_options} and 
atmospheric neutrino experiments \cite{sk,hri} can also determine the hierarchy. 
The measurement of $\dcp$ is 
difficult but possible, thanks to the non-zero value of $\tjl$. This requires 
very intense beams at short baselines \cite{campagne}. In this work, we 
concentrate on the precision measurement of the 
atmospheric mixing angle. 

$\tkl$ is the largest of the mixing angles in the leptonic and quark sectors. 
Its near-maximal value 
is indicative of a symmetry of nature in the $\mu-\tau$ sector \cite{magic1,mutausym}. 
In many theoretical models, the deviation of $\tkl$ from maximality is related 
to the deviation of $\tjl$ from zero \cite{mutausym}. 
Thus, the precision measurement of this angle, and determination of its octant can play 
an important role in constructing new models of physics. 

In this paper, we discuss the measurement of $\tkl$ and its octant, particularly in 
light of moderately 
large $\tjl$, which gives rise to three-flavour effects. In 
Ref.~\cite{parke_defn,degouvea_defn}, 
the authors had shown that due to three-flavour effects, the mass-squared difference 
$\mm$ measured in muon disappearance experiments is not $\mlj$, but a linear 
combination of $\mlj$ and $\mkj$. The same calculation also indicates that the mixing 
angle $\tm$ measured in these experiments is not the same as $\tkl$. 
While this result 
has been seen in the literature before (in Ref.~\cite{old_spurioust23}) and more recently 
in Ref.~\cite{schwetz_nufit}, we present a detailed study of this effect in this paper. 
We pay particular attention to the effect of choosing the `wrong' definition 
($\tkl=\tm$) in analyses. In Section II, we have outlined the calculation that 
indicates the relation between $\mm,\tm$ and $\mlj,\tkl$. We have also found 
analytic expressions relating deviations from maximality in the two octants. In this
work, our aim is to highlight a physics point, rather than study the capability of
any particular experiment. However, in order to make our point clearer, we have 
presented the results of some simulations, in Section III. Finally, in Section IV, 
we have summarized our findings. 

\section{Calculations}

The $\pmumu$ oscillation probability in the three-flavour scenario 
(ignoring matter effects) is given by
\begin{eqnarray}
\pmumu = 1 &-& 4 |U_{\mu3}|^2 |U_{\mu1}|^2 \sin^2\hatmlj \nonumber \\
 &-& 4 |U_{\mu3}|^2 |U_{\mu2}|^2 \sin^2\hatmlk \nonumber \\
 &-& 4 |U_{\mu2}|^2 |U_{\mu1}|^2 \sin^2\hatmkj ~,
 \label{pmm3flav}
\end{eqnarray}
where we have used the shorthand notation $\hat{\Delta}_{ij}  = \Delta_{ij}L/4E$. 
Here, $U_{\alpha i}$ are the elements of the $3\times3$ PMNS matrix.
This probability is sensitive to all six standard oscillation parameters - 
three mixing angles, two mass-squared differences and the CP phase. 
In interpreting the result of oscillation data in terms of two-flavour 
oscillations, we attempt to express the probability in the simple form 
\be
 \pmumu = 1 - \sin^2 2\tm \sin^2 \hatmm 
\ee
that involves only two parameters. Recasting the full six-parameter formula as a 
simple two-parameter formula results in a non-trivial relation between $\mm$ and $\mlj$, 
and between $\tm$ and $\tkl$. 
In this calculation, we will consistently retain only terms upto linear order 
in $\mkj$ (since $\mkj<<\mlj$). Ignoring the last term in 
Eq.~\ref{pmm3flav}, we have 
\be
\frac{1-\pmumu}{4} = |U_{\mu3}|^2 |U_{\mu1}|^2 \sin^2\hatmlj + 
|U_{\mu3}|^2 |U_{\mu2}|^2 \sin^2\hatmlk ~.
\ee
Here we introduce the notation
\[
 u_1 = \frac{|U_{\mu1}|^2}{|U_{\mu1}|^2 + |U_{\mu2}|^2} \qquad,\qquad 
u_2 = \frac{|U_{\mu2}|^2}{|U_{\mu1}|^2 + |U_{\mu2}|^2} ~.
\]
Note that $u_1+u_2 = 1$. This notation lets us write 
\be
\frac{1-\pmumu}{4} = |U_{\mu3}|^2 \left( |U_{\mu1}|^2 + |U_{\mu2}|^2 \right)
\left( u_1 \sin^2\hatmlj + u_2 \sin^2\hatmlk \right) ~.
\ee
Using the fact that $\mlk = \mlj-\mkj$ and ignoring the term that is 
quadratic in $\mkj$, a little algebra gives us 
\be
\frac{1-\pmumu}{4} \approx \frac{1}{2} |U_{\mu3}|^2 \left( |U_{\mu1}|^2 + |U_{\mu2}|^2 \right) 
\left[ 1 - \cos 2\hatmlj - 2u_2\hatmkj\sin2\hatmlj \right] ~.
\ee
We rewrite this as 
\begin{eqnarray}
\frac{1-\pmumu}{4} &=& \frac{1}{2} |U_{\mu3}|^2 \left( |U_{\mu1}|^2 + |U_{\mu2}|^2 \right) 
\nonumber \\
&& \left[ 1 - \left( \cos\beta \cos2\hatmlj + \sin\beta \sin2\hatmlj \right) 
\sqrt{1+4 u_2^2 \hatmkj^2} \right] \nonumber \\
&=& \frac{1}{2} |U_{\mu3}|^2 \left( |U_{\mu1}|^2 + |U_{\mu2}|^2 \right) 
\left[ 1 - \cos(2\hatmlj - \beta) \sqrt{1+4 u_2^2 \hatmkj^2} \right] ~.
\end{eqnarray}
where
\[
 \cos\beta = \frac{1}{\sqrt{1+4 u_2^2 \hatmkj^2}} \qquad,\qquad 
\sin\beta = \frac{2 u_2 \hatmkj}{\sqrt{1+4 u_2^2 \hatmkj^2}} ~.
\]
We ignore the quadratic term in the square root, and we note that 
$\beta = \tan^{-1} (2u_2\hatmkj) \approx 2u_2 \hatmkj$. This gives us our final 
result
\be
\pmumu \approx 1 - 4 |U_{\mu3}|^2 \left( 1 - |U_{\mu3}|^2 \right) 
\sin^2 (\hatmlj-u_2 \hatmkj) ~.
\ee
On comparing with the two-flavour formula, we can make the association 
\[
\mlj = \mm + u_2 \mkj ~,
\]
which is the result expressed in Ref.~\cite{parke_defn}. Moreover, we also find that 
\begin{eqnarray}
 \sin^2 2\tm &=& 4 |U_{\mu3}|^2 \left( 1 - |U_{\mu3}|^2 \right) \nonumber \\
  &=& 4 \cos^2\tjl \sin^2\tkl (1 - \cos^2\tjl \sin^2\tkl) ~. \nonumber
\end{eqnarray}
This means that 
\be
\skl = \sin\tm / \cos\tjl \qquad \textrm{or} \qquad \skl = \sin(90^\circ-\tm) / \cos\tjl ~.
\ee

In other words, given a value of $\sskm$, there are two degenerate allowed values of 
$\tm$: $\tmlo$ (in the lower octant) and $\tmho$ (in the higher octant). 
These are related by 
\be
\tmlo = 90^\circ - \tmho ~.
\ee
Then, the corresponding values of $\tkl$ are given by
\be
\sin\tkllo = \frac{\sin\tmlo}{\cos\tjl} \qquad;\qquad \sin\tklho = 
\frac{\sin\tmho}{\cos\tjl} ~.
\label{rightdefn}
\ee
Thus, if $\tjl=0$, we have simply $\tkllo = \tmlo$ and $\tklho = \tmho$. 
However, for non-zero $\tjl$, the relation between $\tkllo$ and $\tklho$
depends on the value of $\tjl$. 

\begin{figure}[h!]
\begin{center}
  \includegraphics[width=0.7\textwidth]{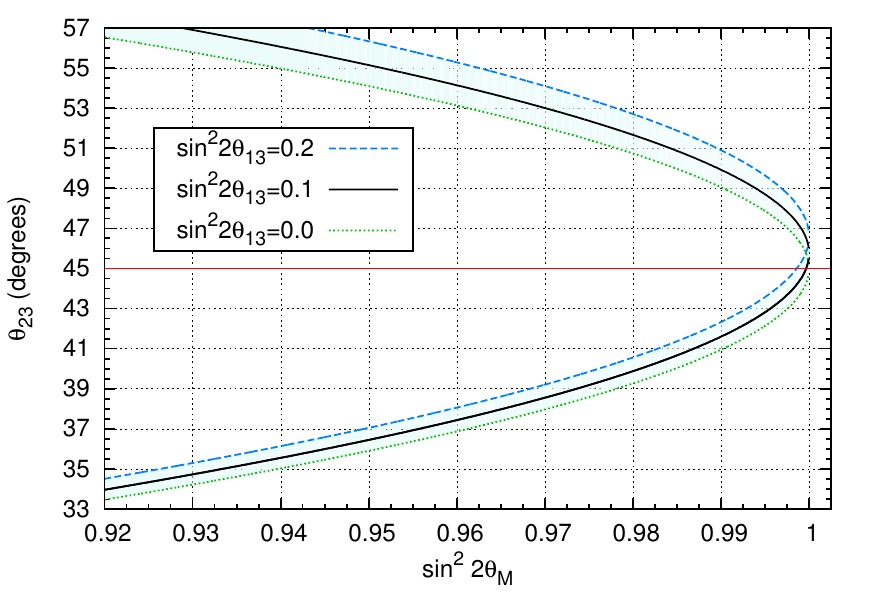}
\end{center}
\vspace{-0.5in}
\caption{\footnotesize
Allowed values for $\tkl$ (y-axis) in degrees, for a given measurement of 
$\sskm$ (x-axis). The dashed (blue) and dotted (green) curves are for the extreme cases 
$\sskjl=0.2$ and $0.0$, respectively. The solid (black) curve is for $\sskjl=0.1$.}
\label{fig:meast23}
\end{figure}

We have shown this feature in Fig.~\ref{fig:meast23}. Along the x-axis, we have 
different values of $\sskm$ as measured using the $\pmumu$ channel. 
For a particular value 
of $\tjl$, each $\tm$ corresponds to two values of $\tkl$. These two values have 
been plotted along the y-axis. Thus given a value of $\tjl$ we have a curve that maps 
one value of $\tm$ to two values of $\tkl$. To show the effect of $\tjl$, we
have varied it in the range $\sskjl \in \left[0.0,0.2\right]$. The solid black 
curve is for $\sskjl=0.1$.
The spread in $\tjl$ gives the shaded band in the figure. 
%%%%%%%%%%%%%%%%%%%%%%%%%%%%%%%%%%%%%%%%%
%%%%%%%%%%%%%%%%%%%%%%%%%%%%%%%%%%%%%%%%%

The main point to be noted here is the asymmetry of the band about the 
$\tkl=45^\circ$ line. When $\sskjl=0$, 
we see that the curve is symmetric about $\tkl=45^\circ$. However, for larger values 
of $\tjl$ such as $\sskjl=0.2$, we lose this symmetry. For instance, 
if $\sskjl= 0.2$ and if the MINOS
measured value of $\sin^2 2\tm= 0.98$, then the values of $\tkl$ in the two octants 
are $40.5^\circ$ and $52.7^\circ$. These two angles are not complementary. 
In other words, it is $\tm$ that goes to $90-\tm$ under the
octant degeneracy; but $\tkllo$ and $\tklho$ are not 
complementary angles - the exact change depends on the value of $\tjl$. Henceforth, 
for convenience, we will refer 
to $\tkllo$ and $\tklho$ corresponding to a given measurement of $\sskm$ as
being \mcomp. 

As an interesting aside, we note that (given $\sskjl=0.1$)
if $44.3^\circ < \tm < 45.7^\circ$, i.e. if 
$\sskm > 0.9993$, then both allowed values of $\tkl$ are greater than $45^\circ$. 
In such a case, there is no ambiguity in the octant of $\tkl$ -- it is necessarily 
in the higher octant. However, current experiments do not have the precision to 
distinguish $\sskm=0.9993$ from $\sskm=1$. Therefore, this point is purely of 
academic interest. 

Hereafter, we assume that $\sskjl=0.1$. When $\sskm=1.0$, i.e. $\tmlo=\tmho=45^\circ$, 
we get $\tkllo=\tklho=45.75^\circ$ from Eq.~\ref{rightdefn}. This is also seen from
Fig.~\ref{fig:meast23}. For a given value of $\sskm \neq 1.0$, the two allowed values of 
$\tm$ have equal and opposite deviations from $45^\circ$, i.e. $\tmlo<45^\circ$ and 
$\tmho>45^\circ$. However, the two \mcomp values of $\tkl$ lie on opposite sides of 
$45.75^\circ$. Since 
\[
 \sin\tmlo = \sin\tkllo \ \cos\tjl ~\textrm{and}
\]
\[
 \sin\tmho = \cos\tmlo = \sin\tklho \ \cos\tjl ~,
\]
one can eliminate $\tmlo$ between these two equations. This gives us 
\be
\sin \tklho = \sqrt{\frac{1}{\cos^2\tjl} - \sin^2 \tkllo} ~,
\label{relateoctants}
\ee
which is a handy equation to switch from one octant to another. For example, if 
$\tkllo=40^\circ$, this equation tells us that the corresponding value of $\tklho$ is
$51.54^\circ$. 

We can try to recast Eq.~\ref{relateoctants} in terms of deviations from maximality, 
rather than in terms of the angles themselves. To this end, we define 
\[
 \delta_{23}^{LO/HO} = \tkl^{LO/HO} - 45^\circ ~.
\]
Assuming small deviations, we can linearize Eq.~\ref{relateoctants}. This gives us 
\be
 \delta_{23}^{HO} = \tan^2 \tjl - \delta_{23}^{LO} \approx 1.5^\circ - \delta_{23}^{LO} ~,
 \label{relatedevs}
\ee
which implies 
\be
 \tklho = 91.5^\circ - \tkllo ~.
 \label{relatelinocts}
\ee
We will use these relations to interpret the results of our simulations. 

In Ref.~\cite{old_spurioust23}, the authors have argued that a precise measurement 
of $\tkl$ is 
difficult in the vicinity of $45^\circ$ because $\sskkl$ attains a maximum here, 
making $\Delta(\sskkl)$ very small. Therefore it is worth discussing whether this 
$\tjl$-effect can have experimentally observable consequences.
In order to directly observe a shift of 
$0.75^\circ$ at $\tm=45^\circ (\tkl=45.75^\circ)$, we need a precision of $0.001$
in our measurement 
of $\sskkl$. This is far beyond our experimental reach. However, 
if $\tmlo=40^\circ (\tkllo=40.63^\circ)$, the precision in 
$\sskkl$ required to observe this shift is around $0.007$, which may be achievable 
at future facilites. Moreover, this effect can be felt indirectly, as an artificially 
enhanced/reduced sensitivity in oscillation experiments. We illustrate this through 
simulations, in the next section. 

\section{Simulations}

As we have mentioned before, our aim is to illustrate the difference between choosing 
the `wrong' definition: $\tkl=\tm$ and the `right' definition: 
$\tkl=\sin^{-1}(\sin\tm/\cos\tjl)$. In order to show the experimentally observable effects 
of this choice, we have simulated the \nova experiment using the GLoBES package 
\cite{nova_tdr,globes1,globes2,globes_nova,messier_xsec,paschos_xsec,Kyoto2012nova}. 
In this section, we discuss the results of our simulations. 

%%%%%%%%%%%%%%%%%%%%%%%%%%%%%%
\begin{figure}[h!]
\begin{center}
  \includegraphics[width=\textwidth]{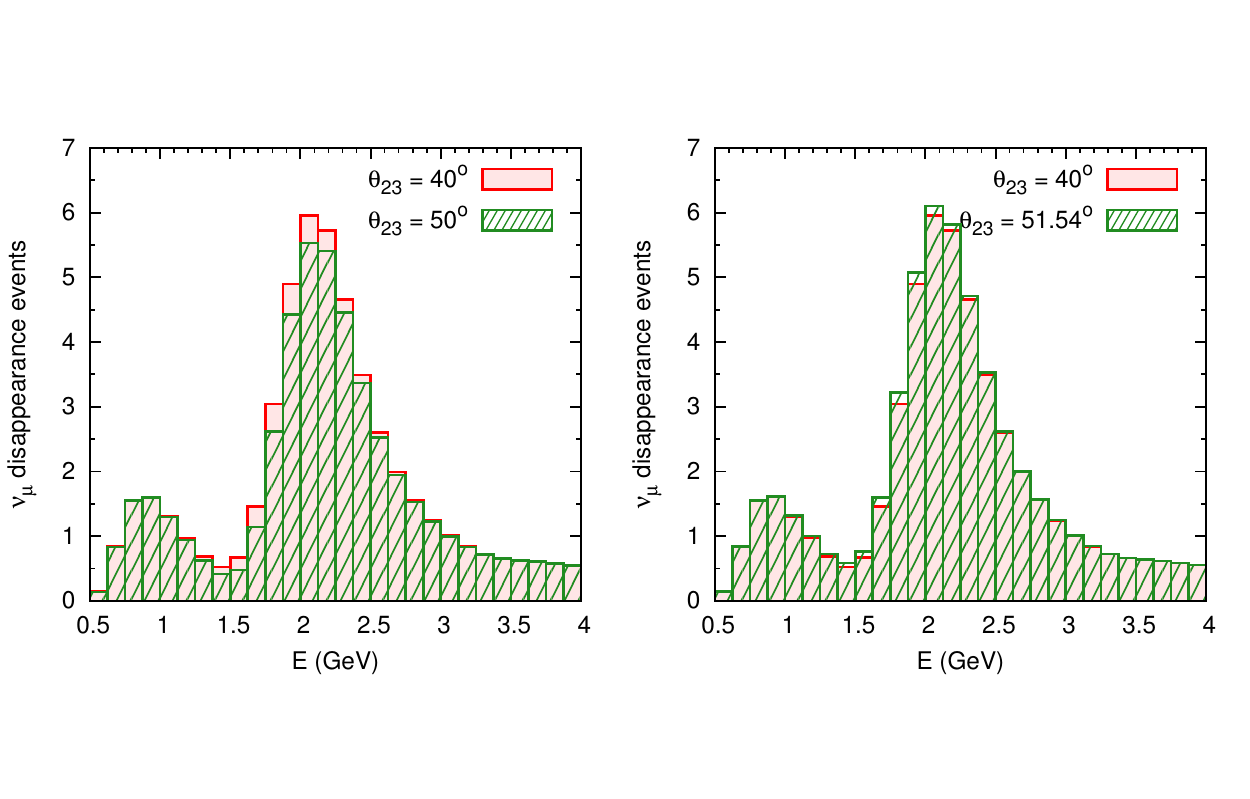}
\end{center}
\vspace{-1in}
\caption{\footnotesize 
$\nu_\mu$ disappearance event rates for $\tkllo=40^\circ$ are shown as a 
solid (red) histogram in both panels. Superimposed on this, as a hatched (green) 
histogram are the event rates for the corresponding $\tklho$ using the `wrong' 
definition (left panel) and `right' definition (right panel). For the `right' 
definition, the event rates show a good match.}
\label{fig:rates40}
\end{figure}
%%%%%%%%%%%%%%%%%%%%%%%%%%%%%%

We have used the standard \nova setup \cite{nova_tdr}, with a 14 kton 
totally active scintillator detector placed 810 km away from the NuMI beam at a 
14 mrad off-axis location. The beam, with a power of 0.7 MW, is assumed to run for 
three years each with neutrinos and antineutrinos. We have taken the energy resolution 
for $\nu_\mu$ to be $0.06\sqrt{E \textrm(GeV)}$ \cite{Kyoto2012nova}. Backgrounds 
from NC events 
have also been taken into account. For the oscillation parameters, we have chosen 
$\sin^2\tjk=0.304$, $\sin^2 2\tjl=0.1$, $\mkj=7.6\times10^{-5} \ \textrm{eV}^2$, 
$\mm=2.4\times10^{-3} \ \textrm{eV}^2$ and $\dcp=0$, unless specified otherwise.

In Fig.~\ref{fig:rates40}, we have plotted the event rates from the muon disappearance 
channel. In the left panel, we have chosen the `wrong' definition of $\tkl$, so that 
$\tkllo=40^\circ$ implies $\tklho=50^\circ$. This gives us a difference in the number 
of events. However, on using the `right' definition ($\tklho=51.54^\circ$), we find 
that the event rates match, as seen in the right panel. 

Having showed the difference due to our choice of $\tkl$ at the event level, we now 
proceed to do so at the level of $\xx$. Throughout our simulations, we have used the 
values specified above as the true values of oscillation parameters. We have varied 
the test values of the parameters in the following ranges: $\sskjl \in [0.07,0.13]$, 
$\tm \in [35^\circ,55^\circ]$, $|\mm| \in [2.05,2.75]\times 10^{-3} \ \textrm{eV}^2$ 
and $\dcp \in [0,2\pi)$. The solar parameters $\mkj$ and $\tjk$ have been kept 
fixed in this analysis, since the effect of their variation is small. The mass 
hierarchy is assumed to be normal. 

%%%%%%%%%%%%%%%%%%%%%%%%%%%
\begin{figure}[h!]
\begin{center}
  \includegraphics[width=\textwidth]{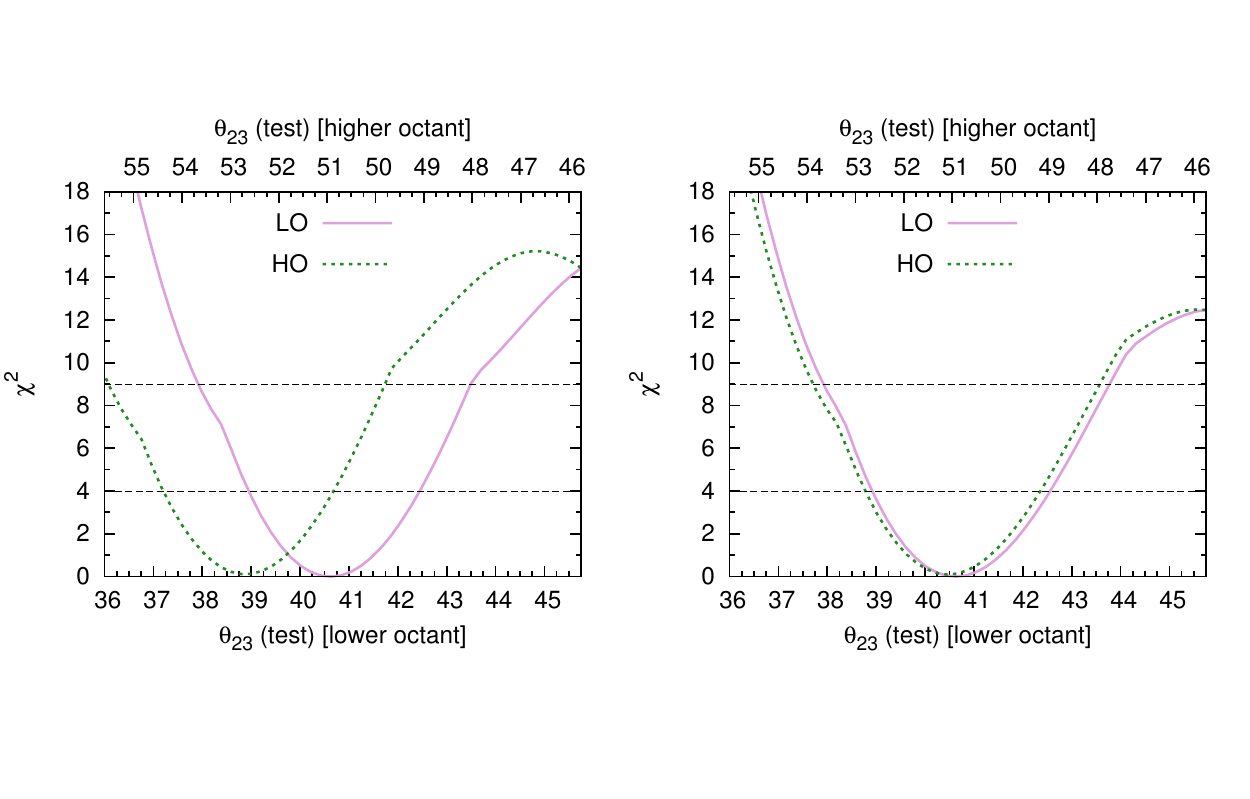}
\end{center}
\vspace{-1in}
\caption{\footnotesize
$\xx$ for sensitivity to the value of $\tkl$ from muon disappearance from \nova. 
True $\tm=40^\circ$. The solid (pink) curve gives $\xx$ for test $\tm$ in the 
lower octant (read the lower x-axis), wile the dotted (green) curve gives 
$\xx$ for test $\tm$ in the higher octant (read the upper x-axis). In the left panel, 
we have chosen the `wrong' definition of $\tkl$ while in the right panel we have chosen 
the `right' definition of $\tkl$.
}
\label{fig:chisq40}
\end{figure}
%%%%%%%%%%%%%%%%%%%%%%%%%%%

In Fig.~\ref{fig:chisq40}, we have plotted the sensitivity 
to $\tkl$ for true $\tm=40^\circ$. As the test value of $\tmlo$ increases in the 
lower (true)
octant from $35^\circ$ to $45^\circ$, $\tkllo$ increases from $35.53^\circ$ to 
$45.75^\circ$. This range is shown on the lower x-axis. Correspondingly, $\tklho$ 
decreases from $55.60^\circ$ to $45.75^\circ$. This range, for the higher 
(false or degenerate) octant, 
is shown on the upper x-axis. The advantage of using these double-axes is that values 
of $\tkl$ along a vertical line are \mcomp. 
We have used the linearized 
relation in Eq.~\ref{relatelinocts} to plot these axes.
For values of test $\tm$ in the lower octant, $\xx$ 
has been plotted as the solid (pink) curve. The lower x-axis should be used to read 
the $\tkl$ values for this curve. For values of test $\tm$ in the higher octant, $\xx$ 
has been plotted as the dotted (green) curve. The upper x-axis should be used to read 
the $\tkl$ values for this curve. 

%%%%%%%%%%%%%%%%%%%%%%%%%%%
\begin{figure}[h!]
\begin{center}
  \includegraphics[width=\textwidth]{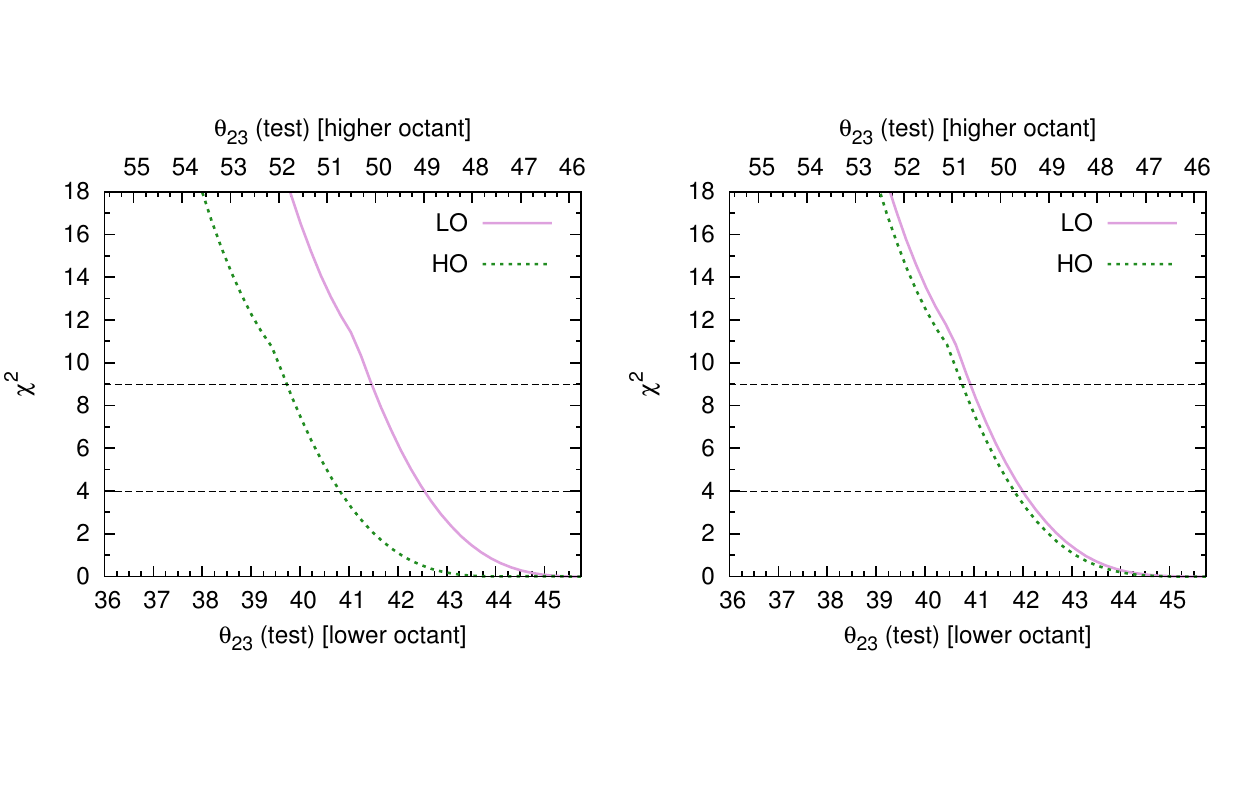}
\end{center}
\vspace{-1in}
\caption{\footnotesize
$\xx$ for sensitivity to the value of $\tkl$ from muon disappearance from \nova. 
True $\tm=45^\circ$. The solid (pink) curve gives $\xx$ for test $\tm$ in the 
lower octant (read the lower x-axis), wile the dotted (green) curve gives 
$\xx$ for test $\tm$ in the higher octant (read the upper x-axis). In the left panel, 
we have chosen the `wrong' definition of $\tkl$ while in the right panel we have chosen 
the `right' definition of $\tkl$.}
\label{fig:chisq45}
\end{figure}
%%%%%%%%%%%%%%%%%%%%%%%%%%%

%%%%%%%%%%%%%%%%%%%%%%%%%%%%%%%
%\begin{figure}[h!]
%\begin{center}
%	\epsfig{file=cont_allowed23_dapp_multi.eps,width=0.9\textwidth}
%\end{center}
%\vspace{-0.75in}
%\caption{\footnotesize
%Allowed test values for $\tkl$ (y-axis) in degrees, for a given true value of 
%$\tkl$ (x-axis). $68\%$, $90\%$ and $95\%$ C.L. contours are also shown. As before, the 
%left(right) panel is plotted using the `wrong(right)' definition of $\tkl$. With the `right' 
%definition, we see the expected behaviour - the false degenerate solution lies along the line 
%$\tklho=91.5^\circ-\tkllo$.}
%\label{fig:allowed23}
%\end{figure}
%%%%%%%%%%%%%%%%%%%%%%%%%%%%%%%
%%%%%%%%%%%%%%%%%%%% 
\begin{figure}[h!]
% \begin{center}
  \begin{tabular}{cc}
  \includegraphics[width=0.5\textwidth]{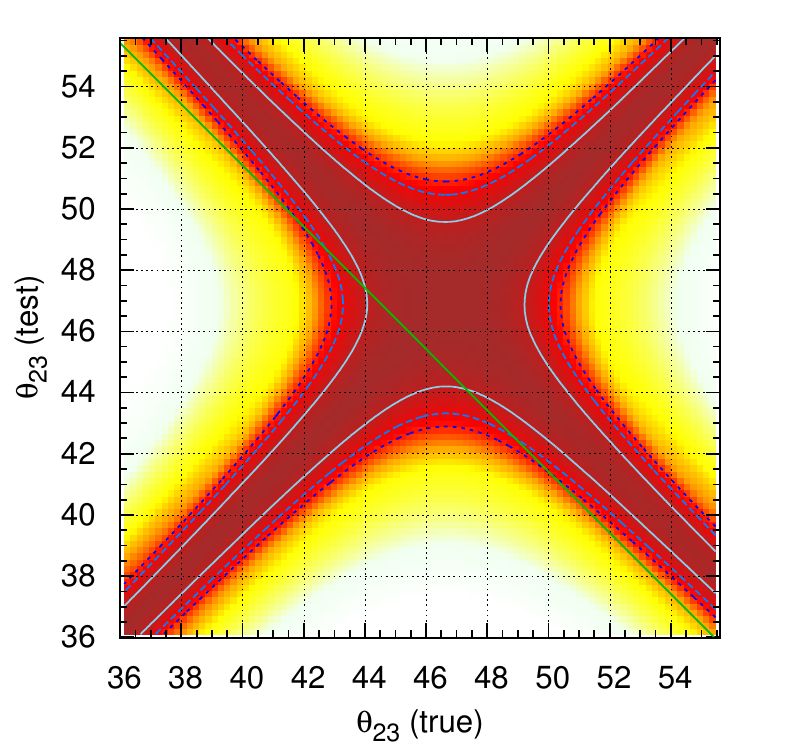}
  \includegraphics[width=0.5\textwidth]{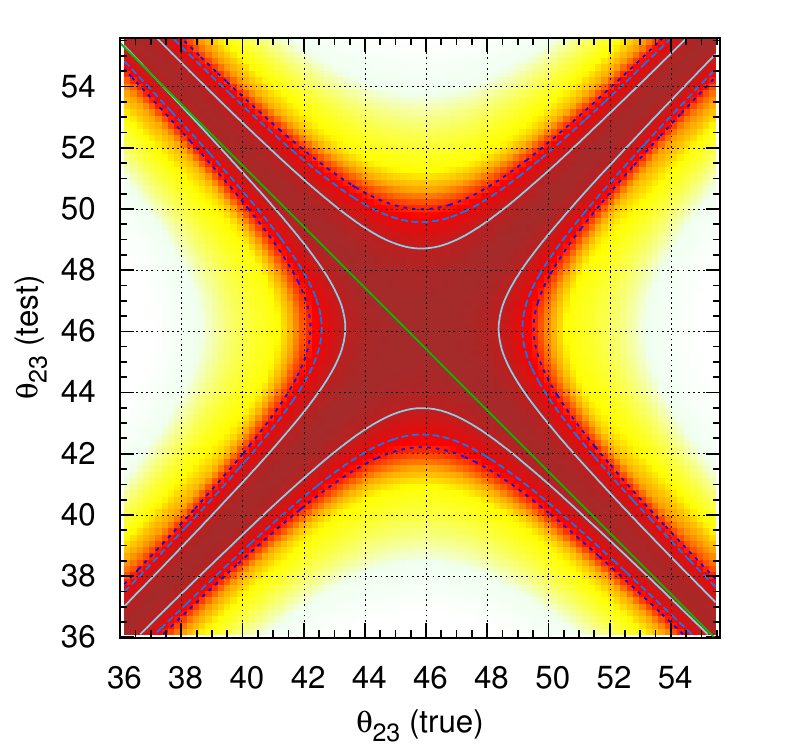}
  \end{tabular}
  \vspace{-0.25in}
  \caption{\footnotesize Allowed test values for $\tkl$ (y-axis) in degrees, for a 
  given true value of $\tkl$ (x-axis). $68\%$, $90\%$ and $95\%$ C.L. contours are 
  also shown. As before, the left(right) panel is plotted using the `wrong(right)' 
  definition of $\tkl$. With the `right' definition, we see the expected behaviour 
  - the false degenerate solution lies along the line $\tklho=91.5^\circ-\tkllo$
  (shown by the straight diagonal (green) line).}

% \end{center}
  \label{fig:allowed23}
\end{figure}
%%%%%%%%%%%%%%%%%%%%%

The following features are visible in Fig.~\ref{fig:chisq40}: (a) For $\tm=40^\circ$, 
the best-fit point is seen at the corresponding value of $\tkl$ which is $40.63^\circ$ 
(b) If we use the `wrong' 
definition of $\tkl$, the minima in the two octants do not appear for \mcomp values of 
$\tkl$. This is seen in the left panel. In the right panel, we have used the `right' 
definition. As a result, we find that the minima coincide. (c) The sensitivities 
in the two octants 
coincide at $\tkl=45.75^\circ$, rather than $\tkl=45^\circ$. This is expected from 
the analytic calculations presented in the previous section. 

Figure~\ref{fig:chisq45} is similar to Fig.~\ref{fig:chisq40}, but with true 
$\tm=45^\circ$. Note that compared to the `right' definition $\xx$, the `wrong' 
definition $\xx$ is more in the lower octant, and less in the higher octant. 
Clearly, using the `wrong' definition gives us an incorrect value of $\xx$. 
Thus, if $\xx$ from the `wrong' definition is used to set a prior on $\tkl$ for 
future experiments, it will 
give a spurious indication of the confidence with which certain $\tkl$ values
are allowed. This also has implications for octant sensitivity studies. Since 
$\tkl$ and $90^\circ-\tkl$ are not \mcomp\!, $\xx(\tkl;90^\circ-\tkl)$ does not 
give the correct octant sensitivity. Therefore, care must be taken to use the `right' 
definition in simulations.

Finally, in Fig.~\ref{fig:allowed23}, we have plotted the allowed values of $\tkl$ for 
all possible true values of $\tkl$ in the range $[36^\circ,55^\circ]$. For each true 
value, the allowed test values can be read from its corresponding vertical line. The 
$68\%$, $90\%$ and $95\%$ C.L. contours are also shown. For a given true value of $\tkl$, 
we find an allowed range of test $\tkl$ in each octant. For the `right' definition, 
we see that the false degenerate solution lies along the line $\tklho=91.5^\circ-\tkllo$ 
(Eq.~\ref{relatelinocts}). But if we choose the 
`wrong' definition, then we get degenerate solutions that are not \mcomp, along with 
incorrect values of $\xx$. 

In this 
paper, we have only presented the results for the case where NH is the true hierarchy, and 
true $\dcp=0$. However, we have verified that these results hold for IH, and for a number of 
values of $\dcp$.

\section{Conclusions and Summary}

In this study, we have discussed the effect of three-flavour mixing on our measurement 
of $\tkl$. We have calculated the relation between the value of $\tkl$ in nature, and $\tm$ 
that is measured in muon disappearance experiments. The difference between these two numbers 
is significant, in light of the measured value of $\tjl$.
Using the `right' definition of 
$\tkl$ (incorporating the $\tjl$-effect), we have found the allowed values of $\tkl$ in 
the two octants, corresponding to a single measured value of $\sskm$. We know that 
complementary values of $\tm$ in the 
two octants are related by $\tmho = 90^\circ - \tmlo$. But for a given measurement of $\sskm$, 
we have found that the corresponding values of $\tkl$ in the two octants are related by 
$\tklho = 91.5^\circ - \tkllo$. We have called this relation $\mu$-complementarity. The 
exact form of this equation comes from the value of $\tjl$. 

Through simulations, we have showed that using the `wrong' definition gives us degenerate 
fits at non-\mcomp values of $\tkl$. Consequently, the muon disappearance analysis can give an 
incorrect prior on $\tkl$. In 
determining octant sensitivity, if the `wrong' definition is used, one can get
an incorrect value of $\xx$. 
Therefore, we advocate the use of the `right' definition (as given 
in Eq.~\ref{rightdefn}) for calculating the oscillation probability in simulations. 
However, priors from 
previous experiments should be added in terms of $\sskm$ -- the quantity that is measured 
in the 
muon disappearance experiments. Our results and conclusions have been found to hold true for 
both hierarchies and for values of $\dcp$ in its entire allowed range.

\begin{acknowledgments}
The author would like to thank Srubabati Goswami and S. Uma Sankar for useful 
discussions, and for a critical reading of the manuscript. 
\end{acknowledgments}

\bibliographystyle{apsrev}
\bibliography{neutosc}

\end{document}